# CALIBRATING *CHEVRON* FOR PREEMPTION

GREGORY M. DICKINSON*

TABLE OF CONTENTS

\*   Associate, Ropes & Gray, LLP; J.D., Harvard Law School; B.S., Houghton College. Thanks to John Manning and Matthew Stephenson for helpful comments on an earlier draft.

## INTRODUCTION

For years now, courts and commentators have struggled to reconcile the presumption against preemption—the interpretive canon that presumes against federal incursion into areas of traditional state sovereignty—with the Court's *Chevron* doctrine, which instructs courts to defer to reasonable agency interpretations of ambiguous federal statutes. Where Congress's preemptive intent is ambiguous, should courts defer to agency interpretations under *Chevron*, or do preemption's federalism implications demand a less deferential approach? Despite numerous opportunities, the Supreme Court has failed to clearly define the level of deference due to preemptive agency interpretations.[1] In some cases the Court appears quite deferential and in others almost entirely nondeferential.

Academic treatment of the Court's jurisprudence has been rightly critical. The Court's unpredictable approach sows uncertainty among regulated parties, the lower courts, and the agencies themselves. As alternatives to the Court's current case-by-case approach, commentators have advocated a variety of more rule-like regimes: universal nondeference, universal *Chevron* deference, and, most commonly, universal *Skidmore* deference.[2] Advocates of across-the-board nondeference point to the lack of political and procedural safeguards protecting states from agency-initiated preemption.³ Those advocating across-the-board *Chevron* deference, on the other hand, point to agencies' technical expertise on preemption questions and the availability of the *Mead* doctrine as a screen to protect the values of federalism where agencies act other than with the force of law.⁴ Finally, a third set of commentators attempts to reconcile these competing approaches by adopting a middling standard of *Skidmore* deference based on the thoroughness and persuasiveness of an agency's judgment in a particular case.⁵

---

    1. *See, e.g.*, Riegel v. Medtronic, Inc., 128 S. Ct. 999, 1010−12 (2008); Watters v. Wachovia Bank, N.A., 550 U.S. 1, 20–21 (2007) (noting but failing definitively to resolve the "academic question" of what deference is owed to an agency's preemption determination).

    2. *See* Nina A. Mendelson, Chevron *and Preemption*, 102 MICH. L. REV. 737, 797−800 (2004) (suggesting a regime of lower, *Skidmore*-style deference as an alternative to the *Chevron* doctrine); Thomas W. Merrill, *Preemption and Institutional Choice*, 102 NW. U. L. REV. 727, 729 (2008) (suggesting significant deference to agency assessments of the need for federal preemption); Catherine M. Sharkey, *Products Liability Preemption: An Institutional Approach*, 76 GEO. WASH. L. REV. 449, 491−98 (2008) (suggesting a *Skidmore*-like regime in which, though courts do not defer to agency *conclusions* regarding preemption, they recognize agencies' superior ability to supply and analyze empirical data relevant to the desirability of a uniform federal regulatory system); Ernest A. Young, *Executive Preemption*, 102 NW. U. L. REV. 869, 871 (2008) (suggesting that the *Chevron* doctrine should give way in the face of the constitutionally grounded presumption against preemption); *The Supreme Court, 2008 Term— Leading Cases*, 123 HARV. L. REV. 262, 271−72 (2009) (dicussing *Wyeth v. Levine* and suggesting that a *Chevron*-based regime would most appropriately recognize agency expertise).

Thus far, none of these approaches have tempted the Court. Instead, the Court continues to apply deference haphazardly from case to case with no clearly articulated reason for its variation. A close study of the cases, however, reveals both why the Court has been reluctant to adopt any of the proposed across-the-board standards of deference and what an appropriate framework for agency deference might look like. The Court's inconsistent decisionmaking stems from its high regard for congressional intent when considering questions that implicate federalism. *Chevron* and the presumption against preemption provide conflicting indicia of congressional intent, and rather than universalize one principle at the expense of the other, the Court has applied deference selectively depending on its case-specific analysis of congressional intent. When the Court thinks it reasonable to presume delegation of preemptive authority, it is quite deferential to agency views. But, when it thinks congressional intent to delegate is unlikely, it accords little deference to preemptive agency interpretations.

Critics of the Court's *Chevron*–preemption jurisprudence correctly note its major flaw—its inconsistency—but they fail to recognize its purpose and benefits. By looking to congressional intent rather than universalizing a sometimes-inapplicable, across-the-board rule, the Court respects congressional intent where it intends to delegate preemptive authority, while protecting state sovereignty where it does not. Of course, the Court's good intentions do not excuse the approach's unpredictability. A superior approach would package the Court's concern for state sovereignty and congressional intent into a predictable and easily administrable bright-line rule.

---

3. *See, e.g.*, Young, *supra* note 2, at 869 ("As the constitutional limits on national action fade into history, the primary remaining safeguards for state autonomy are political, stemming from the representation of the states in Congress, and procedural, arising from the sheer difficulty of navigating the federal legislative process. These safeguards have little purchase on executive action.").

4. *See, e.g.*, *Leading Cases*, *supra* note 2, at 272 ("[W]hile there are enduring concerns with respect to agency interpretation of preemption questions, the traditional *Chevron/Mead* deference framework can address these concerns, with no need for a singular approach for preemption questions. Bringing the doctrine in this area in line with the overall agency deference approach promises . . . to take advantage of agency interpretive strengths . . . .").

5. *See, e.g.*, Mendelson, *supra* note 2, at 797–800 (suggesting that although full *Chevron*-style deference is inappropriate in the preemption context, agencies' expertise in interpreting and administering complex regulatory statutes counsels in favor of *Skidmore* deference); Sharkey, *supra* note 2, at 491–98 (suggesting a *Skidmore*-like regime because of agencies' peculiar competency to interpret the complexity of the statutes that they administer).

The Court's existing doctrinal distinction between express and implied preemption points to a possible solution. In express preemption cases, the Court does not need to enforce federalism values through the presumption against preemption because Congress has spoken clearly in favor of displacing state law. And if the scope of preemption is ambiguous, *Chevron*'s presumption of delegation through ambiguity to agency expertise is entirely reasonable. Agencies are quite competent to decide the proper scope of preemption once Congress has duly authorized it. On the other hand, where Congress has not spoken clearly through an express preemption clause, and the question is whether there is to be any preemption at all, *Chevron*'s rationale is particularly weak. Agencies are least competent when considering unbounded questions of federal–state power allocation, and Congress is unlikely to delegate authority of this sort.

Given the waxing and waning force of *Chevron*'s rationale across cases, the Court should adopt a rule of variable deference that accords full *Chevron*-style deference to agency interpretations of ambiguously broad express preemption clauses and withholds deference altogether where Congress is silent regarding preemption. Such a rule, unlike any of the proposed across-the-board regimes, would recognize the factors that underlie the Court's unpredictable case-by-case approach—respect for state sovereignty and congressional intent—while providing the rule-like certainty demanded by the Court's critics.

Part I presents a brief overview of current preemption law and the conflicting rationales underlying the *Chevron* doctrine and the presumption against preemption. Part II closely examines the Court's recent case law and concludes that the Court's inconsistency stems from its direct scrutiny of legislative intent. Because of its effort to respect federalism values while sidestepping the conflicting canons, the Court's analysis has descended to an unpredictable case-by-case search for congressional intent. Part II also explains why no regime of uniform, across-the-board deference can adequately account for the Court's concerns: *Chevron*'s presumption of congressional delegation applies in some cases more than others. Finally, Part III presents a framework for deference in *Chevron*–preemption cases that conditions deference, in rule-like fashion, on the presence of an express preemption clause—accounting for congressional intent to delegate while ensuring predictability.

I. THE *CHEVRON*–PREEMPTION CLASH

To evaluate the Court's handling of the conflict between *Chevron* and the presumption against preemption, it is useful first to pause and consider the rationales underlying those doctrines and the current state of preemption law generally.

*A. Federal Preemption of State Law*

Congress's power of preemption, rooted in the Supremacy Clause of the Constitution,[6] permits federal law to trump state law where it is undesirable or impossible for two independent legal regimes to coexist. The Supreme Court has recognized two primary categories of preemption: express and implied.[7] Express preemption occurs where a federal statute expressly withdraws regulatory power over a particular area of law from the states.[8] Express preemption doctrine therefore involves the difficult but familiar judicial task of determining the intended preemptive reach of statutory language.[9] Implied preemption is subdivided into two types: field preemption and conflict preemption.[10] Field preemption occurs where a federal regulatory regime is so pervasive as to imply that Congress intended to occupy an entire field of the law, leaving no room for states to supplement that federal regulation.[11] Similarly, but on a smaller scale, conflict preemption occurs where, though Congress has demonstrated no intent to occupy an entire field of law, federal law conflicts with a particular state law.[12] This conflict may take either of two forms: First, state law will be preempted "where it is impossible for a private party to comply with both state and federal law."[13] Second, state law will also be preempted where, though it is not literally impossible to comply with both state and federal law, state law "stands as an obstacle to the accomplishment and execution of the full purposes and objectives of Congress."[14] This taxonomy of preemption yields four fundamental varieties: express preemption, field preemption, impossibility preemption, and obstacle preemption. Though this Article will not dwell on the nuanced distinctions among the doctrines, it is important at the outset to recognize the basic distinction between express and implied varieties of preemption. The Court is much more skeptical of implied preemption claims than it is of express preemption claims,[15] and that skepticism factors heavily in its treatment of agency determinations for or against preemption.[16]

---

6. U.S. CONST. art. VI, cl. 2 ("This Constitution, and the Laws of the United States . . . shall be the supreme Law of the Land; and the Judges in every State shall be bound thereby, any Thing in the Constitution or Laws of any State to the Contrary notwithstanding.").

7. *See* Gade v. Nat'l Solid Wastes Mgmt. Ass'n, 505 U.S. 88, 98 (1992) ("Pre-emption may be either expressed or implied, and 'is compelled whether Congress' command is explicitly stated in the statute's language or implicitly contained in its structure and purpose.'" (quoting Jones v. Rath Packing Co., 430 U.S. 519, 525 (1977))).

8. Caleb Nelson, *Preemption*, 86 VA. L. REV. 225, 226–27 (2000).

9. *See id.* (explaining that judges faced with an express preemption clause must determine both the meaning of the clause and whether the Constitution allows Congress to forbid the states from exercising the powers in question); Daniel E. Troy & Rebecca K. Wood, *Federal Preemption at the Supreme Court*, 2007–2008 CATO SUP. CT. REV. 257, 258 (2008).

10. *Gade*, 505 U.S. at 98.

11. *See* Cipollone v. Liggett Grp., Inc., 505 U.S. 504, 516 (1992) ("In the absence of an express congressional command, state law is pre-empted if that law actually conflicts with federal law, or if federal law so thoroughly occupies a legislative field 'as to make reasonable the inference that Congress left no room for the States to supplement it.'" (quoting Rice v. Santa Fe Elevator Corp., 331 U.S. 218, 230 (1947)); English v. Gen. Elec. Co., 496 U.S. 72, 79 (1990) ("[I]n the absence of explicit statutory language, state law is pre-empted where it regulates conduct in a field that Congress intended the Federal Government to occupy exclusively."); *see also* Nelson, *supra* note 8, at 227.

### B. The Presumption Against Preemption

Regardless of the particular preemption doctrine involved, preemption questions are enormously important. The extent to which federal law displaces state law determines the legal regime or regimes under which particular cases will be decided and, more broadly, the balance of power between the states and the federal government.[17] Overpreemption threatens to extinguish the states' traditional sovereign roles as checks on federal power and guarantors of individual rights,[18] while underpreemption

---

12. *See* Nelson, *supra* note 8, at 228.
13. Crosby v. Nat'l Foreign Trade Council, 530 U.S. 363, 372–73 (2000); *see, e.g.*, Fla. Lime & Avocado Growers, Inc. v. Paul, 373 U.S. 132, 142–43 (1963) (upholding a California statute where dual compliance with both state and federal law was possible).
14. Hillsborough Cnty. v. Automated Med. Labs., Inc., 471 U.S. 707, 713 (1985) (quoting Hines v. Davidowitz, 312 U.S. 52, 67 (1941)).
15. *See id.* at 714 (noting that a defendant advancing an argument of implied preemption "faces an uphill battle"); Mary J. Davis, *The Battle over Implied Preemption: Products Liability and the FDA*, 48 B.C. L. REV. 1089, 1132–33 (2007) (arguing that the Supreme Court demands strong clear evidence of implied conflict because it is a weak substitute for congressional intent).
16. *See infra* Part II.B.
17. Nelson, *supra* note 8, at 225–26.
18. *See* Betsy J. Grey, *Make Congress Speak Clearly: Federal Preemption of State Tort Remedies*, 77 B.U. L. REV. 559, 613–18 (1997) (arguing that overpreemption threatens states control over tort law); Roderick M. Hills, Jr., *Against Preemption: How Federalism Can Improve the National Legislative Process*, 82 N.Y.U. L. REV. 1, 16–18 (2007) (arguing that Congress has institutional tendencies to defer politically sensitive issues to bureaucratic resolution and that less threatens the efficiency provided by uniform federal regulatory schemes.[19] And of course, in any given case, the parties will have their own self-interested views on the proper law to apply as well.

Recognizing the delicacy and importance of preemption questions, the Supreme Court has generally applied a presumption against preemption of state law, requiring from Congress a clear statement of intent to preempt before it is willing to find state law preempted by a federal statute.[20] The Court's classic statement of the principle is found in *Rice v. Santa Fe Elevator Corp.*[21]: "[W]e start with the assumption that the historic police powers of the States [are] not to be superseded by [a] Federal Act unless that was the clear and manifest purpose of Congress."[22] This presumption, which effectively forces congressional deliberation by requiring an explicit preemption clause or its equivalent, can be justified under a number of theories.

First, the presumption represents an embrace of federalism values and a reluctance to risk incidental interference with state sovereignty.[23] By forcing Congress to speak clearly, the Court protects parallel state legal regimes from federal incursion and thereby promotes all of the traditionally recited advantages of divided sovereignty.[24] Second, the presumption against preemption also reflects an empirical assumption regarding legislative intent. Given our nation's traditional system of limited federal government and respect for state autonomy, courts may be justified in presuming, absent clear evidence to the contrary, that federal legislators do not intend their efforts to displace existing state law.[25]

---

preemption would permit states to force issues onto the congressional agenda through state legislative efforts); S. Candice Hoke, *Preemption Pathologies and Civic Republican Values*, 71 B.U. L. REV. 685, 710–14 (1991) (arguing that overpreemption threatens public participation in state political processes).

19. *See* Alan Schwartz, *Statutory Interpretation, Capture, and Tort Law: The Regulatory Compliance Defense*, 2 AM. L. & ECON. REV. 1, 20–22 (2000) (arguing that inefficiencies result from nonuniform state safety standards).

20. Mendelson, *supra* note 2, at 752.

21. 331 U.S. 218 (1947).

22. *Id*. at 230.

23. *See* Mendelson, *supra* note 2, at 756 (citing as an example the *Medtronic* Court's description of federal preemption as a "serious intrusion into state sovereignty" (quoting Medtronic, Inc. v. Lohr, 518 U.S. 470, 488 (1996))).

24. *See id.* at 756–57 & n.76 (collecting sources and including government responsiveness, promotion of self-governance, efficiency, and interstate competition for citizens in a list of federalism's traditionally recited values).

25. *Id.* at 755.

### C. *The* Chevron *Doctrine*

Like the presumption against preemption, the *Chevron*[26] doctrine also relies on a presumption regarding congressional intent to interpret difficult statutory language. The doctrine presumes that Congress would prefer agencies to resolve ambiguities in the statutes they administer and so directs courts to defer to an agency's reasonable interpretation of an ambiguous statute. Courts first apply the "traditional tools of statutory construction"[27] to determine "whether Congress has directly spoken to the precise question at issue."[28] If the statutory language is ambiguous, however, courts infer congressional intent to delegate interpretive authority to the agency and defer to the agency's construction as long as it is reasonable.[29]

The inference of intent to delegate is, in many instances, quite reasonable. Congress is a body of generalists with no particular expertise other than lawmaking itself. When drafting or updating the organic statutes underlying complex regulatory regimes, Congress is predictably eager to shift responsibility for technical policy minutiae to experts in executive agencies.[30] Congress may also have other reasons for granting decisionmaking authority to agencies. By enacting skeletal statutes and relying on agencies to fill in the details, Congress is able to take credit for broad initiatives while avoiding blame for more detailed and sometimes controversial policy choices.[31] Thus, statutory ambiguity on a particular question may be a sign that Congress preferred to use imprecise language and thereby delegate ultimate interpretive authority to the agency responsible for administering the statute. The *Chevron* doctrine recognizes this principle and incorporates it into an interpretive canon of deference to agencies' reasonable interpretations of ambiguous statutes.

---

26. Chevron U.S.A. Inc. v. Natural Res. Def. Council, Inc., 467 U.S. 837 (1984).

27. *Id.* at 843 n.9.

28. *Id.* at 842.

29. *Id.* at 843–44 (distinguishing explicit and implicit delegations of rulemaking authority and directing that courts defer to the reasonable interpretation of the agency in ambiguous cases).

30. David Epstein & Sharyn O'Halloran, *The Nondelegation Doctrine and the Separation of Powers: A Political Science Approach*, 20 CARDOZO L. REV. 947, 966–67 (1999) ( "As policy becomes more complex, Congress will rationally rely more on the executive branch to fill in policy details. . . . The first and most obvious reason is that the executive branch is filled (or can be filled) with policy experts who can run tests and experiments, gather data, and otherwise determine the wisest course of policy, much more so than can 535 members of Congress and their staff.").

31. *See* DAVID SCHOENBROD, POWER WITHOUT RESPONSIBILITY: HOW CONGRESS ABUSES THE PEOPLE THROUGH DELEGATION 99–106 (1993) (describing the political cover that broad delegation provides to Congress); Theodore J. Lowi, *Two Roads to Serfdom: Liberalism, Conservation and Administrative Power*, 36 AM. U. L. REV. 295, 296 (1987) ("[T]he delegation of broad and undefined discretionary power from the legislature to the executive branch deranges virtually all constitutional relationships and prevents attainment of the constitutional goals of limitation on power, substantive calculability, and procedural calculability.").

### D. Indeterminacy and Incompatibility: Canons as Rules of Thumb

Though intuitive and quite helpful, both the presumption against preemption and the *Chevron* doctrine are, like all canons of construction, frequently fallible generalizations. They are useful "rules of thumb" but do not always produce the correct result.[32] One can imagine, for instance, scenarios in which Congress does not explicitly state its intent to preempt state law but where its intent to do so is so clear that it would be foolish for a court to allow the presumption against preemption to determinatively affect its interpretation. Indeed, entire branches of preemption doctrine have been built around such cases:

> Even in the absence of an express preemption clause, the Court sometimes is willing to conclude that a federal statute wholly occupies a particular field and withdraws state lawmaking power over that field. The Court has indicated that a federal regulatory scheme may be "so pervasive" as to imply "that Congress left no room for the States to supplement it." . . . In essence, judges who infer such "field" preemption are reading an implicit preemption clause into the federal statute . . . .[33]

Where congressional intent to preempt is clear, that intent wins out despite the general presumption in favor of state sovereignty.

One can imagine similar exceptions to the *Chevron* doctrine. Even where statutory language is imprecise and an agency presents a reasonable construction of that language, deference may be inappropriate if an inference of congressional intent to delegate would be unreasonable. As with the presumption against preemption, the *Chevron* doctrine's rule of thumb can be overcome by sufficiently strong evidence that the canon will incorrectly discern congressional intent in a case or set of cases.[34]

---

32. *See* Varity Corp. v. Howe, 516 U.S. 489, 511 (1996) ("Canons of construction . . . are simply rules of thumb which will sometimes help courts determine the meaning of legislation." (internal quotation marks omitted)); *see also* James J. Brudney & Corey Ditslear, *Canons of Construction and the Elusive Quest for Neutral Reasoning*, 58 VAND. L. REV. 1, 7–14, 29–69 (2005) (providing an overview of recent academic treatment of the canons and an empirical assessment of their use by modern courts).

33. Nelson, *supra* note 8, at 227 (quoting Rice v. Santa Fe Elevator Corp., 331 U.S. 218, 230 (1947)).

34. *See* Thomas W. Merrill & Kristin E. Hickman, Chevron*'s Domain*, 89 GEO. L.J. 833, 872 (2001) (arguing that, because *Chevron* should only apply in cases where Congress intends it to apply, it is important to determine whether Congress would want agencies to have primary interpretational authority).

Accordingly, in a series of post-*Chevron* cases the Court has gradually softened *Chevron*'s rule of deference, carving out exceptions where congressional intent to delegate seems unlikely. In *Adams Fruit Co. v. Barrett*[35] the Court hinted in dicta that *Chevron*-style deference might be inappropriate if Congress did not intend to delegate authority to decide a particular interpretive question.[36] Ten years later, in *Christensen v. Harris County*,[37] the Court declined to extend deference to an agency interpretation contained in an opinion letter rather than in a formal regulation.[38] Citing lack of formality and force of law, the Court appears to have been motivated by a concern that Congress would not have intended to delegate authority to the agency to make important interpretive decisions in such an informal way.[39]

The pattern of exception carving culminated in *United States v. Mead Corp.*[40] where the Court transformed *Chevron*'s hard-and-fast rule of deference to agency interpretations to a more context-specific inquiry into congressional intent to delegate. Similar to *Christensen*, the *Mead* Court declined to extend deference to an agency interpretation contained within an informal agency tariff classification rather than a formal regulation.[41] The Court pushed *Christensen*'s logic a step further, however, reasoning that

---

35. 494 U.S. 638 (1990).

36. In *Adams Fruit Co.*, a group of injured migrant farm workers, who had already received workers' compensation benefits under Florida state law, brought a claim for further benefits under the motor vehicle safety provisions of the Migrant and Seasonal Agricultural Worker Protection Act (AWPA). *Id.* at 640–41. Their employer, Adams Fruit Company, argued that the Court should defer to the Department of Labor's position that where state workers' compensation is available, it should serve as the exclusive remedy. *Id.* at 649–50. The Court rejected this contention, finding the statutory language to unambiguously support the farm workers' position. *Id.* at 650–51. Moreover, the Court reasoned that even were the language ambiguous, Congress had established the Judiciary and not the Department of Labor as the adjudicator of private actions arising under the statute:

> Congress clearly envisioned, indeed expressly mandated, a role for the Department of Labor in administering the statute by requiring the Secretary to promulgate *standards* implementing AWPA's motor vehicle provisions. This delegation, however, does not empower the Secretary to regulate the scope of the judicial power vested by the statute. Although agency determinations within the scope of delegated authority are entitled to deference, it is fundamental that an agency may not bootstrap itself into an area in which it has no jurisdiction.

*Id.* at 650 (internal quotation marks omitted) (citations omitted).

37. 529 U.S. 576 (2000).

38. *Id.* at 586–87.

39. *Id.* at 587 (stating that interpretations contained in opinion letters do not merit the *Chevron*-style deference they otherwise would if made pursuant to notice-and-comment rulemaking).

40. 533 U.S. 218 (2001).

41. *Id.* at 231–32.

even an agency interpretation embodied in a formal regulation would not *necessarily* be entitled to deference.[42] Such formality is a "very good indicator" of congressional intent to delegate but is not alone dispositive.[43] In every instance the Court must ascertain whether the *Chevron* presumption of intent to delegate is reasonable or whether, given the unique circumstances of the case, the doctrine yields an incorrect picture of congressional intent.

Thus, neither the presumption against preemption nor the *Chevron* doctrine of deference erects an unyielding, across-the-board rule. Both interpretive principles are subject to qualification where the Court finds their underlying rationales to be inapplicable to a particular category of cases.[44] The difficulty comes, of course, in delineating areas of inapplicability in a way that is principled and rule-like enough to preserve the utility of the general rule. One might, for instance, along with Justice Scalia, question whether the *Mead* exception to *Chevron* has been framed so loosely as to eviscerate the rule altogether, leaving only a naked inspection of intent to delegate.[45] Such objections raise important questions of proper exception drawing but do not undermine the enterprise itself.[46] Passing over questions of the appropriateness of this or that particular exception, the concern here is simply to illustrate the defeasibility of the canons in the presence of sufficiently strong evidence of contrary congressional intent.

---

42. *Id.* at 229–30. Neither does a lack of formality necessarily preclude *Chevron*-style deference. Post-*Mead*, formality is neither necessary nor sufficient to support judicial deference.

43. *Id.* at 229.

44. The *Christensen–Mead* line is the Court's most direct effort to curtail the scope of *Chevron*, but it has engaged in other instances of exception making as well. *See* Lisa Schultz Bressman, Chevron*'s Mistake*, 58 DUKE L.J. 549, 550–59, 589–602 (discussing two additional cases that eschew traditional *Chevron* analysis where its result would be inconsistent with a theory of congressional delegation: *Zuni Public School District No. 89 v. Department of Education*, 550 U.S. 81 (2007), and *Gonzales v. Oregon*, 546 U.S. 243 (2006)).

45. *See Mead*, 533 U.S. at 240–41 (Scalia, J., dissenting) ("Today the Court collapses [the *Chevron*] doctrine, announcing instead a presumption that agency discretion does not exist unless the statute, expressly or impliedly, says so. . . . The Court has largely replaced *Chevron*, in other words, with that test most beloved by a court unwilling to be held to rules (and most feared by litigants who want to know what to expect): th'ol' 'totality of the circumstances' test.").

46. Courts are in agreement, for instance, that agencies' views on the proper interpretation of the Administrative Procedure Act (APA) are not entitled to *Chevron* deference because no particular agency is assigned a special role in construing that statute. *See* Metro. Stevedore Co. v. Rambo, 521 U.S. 121, 137 n.9 (1997) (noting that *Chevron* deference to an agency interpretation of the APA's burden of proof provision would be inappropriate); *cf.* Rapaport v. Office of Thrift Supervision, 59 F.3d 212, 216 (D.C. Cir. 1995) (determining *Chevron*-style deference to be inappropriate where multiple agencies are responsible for administering a single statute).

In addition to the possibility that in a particular category of cases a canon's rationale will be undercut by stronger, contradictory evidence of congressional intent, interpretive canons are indeterminate in another way as well: often two canons will come into conflict, each pointing toward an opposing result. This possibility of conflict and consequent indeterminacy is amply demonstrated by Karl Llewellyn who, in his classic critique of the canons, presented twenty-eight canons, each side-by-side with its respective countercanon.[47] To say that the canons are indeterminate, however, is not to disclaim their usefulness as interpretive tools. A rule of thumb is useful not for its perfect accuracy but for its broad applicability and ease of application. Canons may be overridden on occasion by superior evidence of statutory meaning without diminishing their utility as tools of interpretation.[48]

The possibility of conflicting canons can be understood as a subset of the more general possibility, discussed in the previous section, that in a particular set of cases superior evidence of congressional intent may defeat a canon. Contrary canons of interpretation are simply one particular way in which congressional intent can be discerned. As in the more general case, the difficulty in resolving canon conflicts lies in determining where each canon should and should not apply (that is, which canon ought to prevail in a particular set of cases) and drawing lines in a rule-like enough fashion that the utility of the general rule is not obliterated.

### E. Defeasibility, Conflict, and Chevron

Notwithstanding interpretive canons' usefulness as tools of statutory construction, they are, as the previous sections have shown, merely rules of thumb, defeasible in the face of sufficiently strong evidence of contrary congressional intent. So, inevitably, when the canons are invoked to solve novel, delicate, or otherwise extraordinary interpretive questions, we begin to question their applicability. We may wonder whether a canon's rationale is applicable at all or, if there is a conflict, which of two canons should win out. These problems pop up with particular severity in the preemption context.

As the modern administrative state has expanded and agencies have become responsible for administering a substantial number of statutes that raise preemption issues, courts have been forced to wrestle with a particularly vexing conflict between the *Chevron* doctrine—which requires deference to an agency's reasonable construction of an ambiguous statute—and the presumption against preemption, which requires a "clear statement" from Congress before a court may conclude that a federal statute preempts state law.[49] Does the federalism-inspired interpretive canon presuming against preemption serve as a "traditional tool[] of statutory construction"[50] resolving ambiguity and obviating the need for *Chevron* deference to agency views? Or does the peculiar competence of agencies within their statutory spheres require deference even regarding such sensitive questions as preemption? And if agency views require deference, should that deference be tempered in light of countervailing federalism concerns? Such questions reflect a deep tension among the *Chevron* doctrine, the presumption against preemption, and their underlying rationales. The question is whether, in the context of preemption, the conflict between *Chevron* and its rationale is severe enough to warrant an exception to the general rule. And if so, how should that exception be framed?

---

47. *See* Karl N. Llewellyn, *Remarks on the Theory of Appellate Decision and the Rules or Canons About How Statutes Are to Be Construed*, 3 VAND. L. REV. 395, 401–06 (1950).
48. *See* ANTONIN SCALIA, A MATTER OF INTERPRETATION: FEDERAL COURTS AND THE LAW 25–27 (Amy Gutman ed., 1997); CASS R. SUNSTEIN, AFTER THE RIGHTS REVOLUTION: RECONCEIVING THE REGULATORY STATE 147–57 (1990).

*Chevron*'s rule of deference is based on the presumption that Congress intends to delegate interpretive authority to agencies to resolve statutory ambiguities. Where such a presumption would be unreasonable, however, *Chevron*'s rationale is undercut. Pointing to the importance of federalism values, agency inexpertness in considering preemption questions, the risk of arbitrary decisionmaking, and the danger of agency self-aggrandizement, Nina Mendelson suggests that preemption questions present just the sort of exceptional circumstance that requires amendment of the general *Chevron* rule.[51] She and other commentators suggest that courts should grant something less than full *Chevron*-style deference to agency determinations of preemption.[52]

As with all instances of exception making, the most difficult question (once it is determined that an exception is indeed necessary) is how best to draw the exception. Line drawing of this sort requires a careful balance. On the one hand, courts must be careful to carve out from the general rule only those instances where the underlying rationale of *Chevron* is inapplicable. On the other hand, overly fine distinctions or case-by-case applications of *Chevron*'s underlying logic risk destroying the utility of *Chevron* as a rule of decision. The remainder of this Article will first examine the

---

49. *See generally* Mendelson, *supra* note 2, at 739–40.
50. *See* Chevron U.S.A. Inc. v. Natural Res. Def. Council, Inc., 467 U.S. 837, 843 n.9 (1984).
51. *See, e.g.*, Mendelson, *supra* note 2, at 779–97.
52. *See supra* notes 2–5 and accompanying text.

Court's halting treatment of the *Chevron*–preemption puzzle and second present a novel, rule-like framework for addressing *Chevron*–preemption cases.

## II. THE COURT'S *CHEVRON*–PREEMPTION JURISPRUDENCE

### A. *Doctrinal Inconsistency and Unpredictable Decisions*

The Court's treatment of *Chevron*'s applicability to questions of federal preemption is notoriously convoluted.[53] Even now, twenty-six years after the *Chevron* decision, it is unclear to what extent *Chevron*'s rule of deference applies in preemption cases. The ambiguity is threefold. First, though the cases generally seem to suggest that full, *Chevron*-style deference is inappropriate in preemption cases,[54] and at least a few Justices are willing to formally renounce the doctrine,[55] the Court has yet to disavow *Chevron*'s applicability in preemption cases. Without a clearly articulated standard, *Chevron*'s applicability must be relitigated in each successive case, and regulated parties are left wondering whether agency determinations will withstand judicial scrutiny.

---

53. *See* Davis, *supra* note 15, at 1093–94 ("The proper weight of an agency's determination of preemptive scope has generated much debate within the Supreme Court and among commentators. The Court has not answered the question of how an agency position affects the operation of implied conflict preemption doctrine, nor has it addressed how the historic primacy of state regulation in the area of health and safety is to be considered in the balance." (footnote omitted)); Paul E. McGreal, *Some* Rice *With Your* Chevron*?: Presumption and Deference in Regulatory Preemption*, 45 CASE W. RES. L. REV. 823, 826 (1995) ("While the Court has spoken on regulatory preemption, it has neither explained nor justified its position. Instead, the Court merely has applied *statutory* preemption rules to *regulatory* preemption cases. To the extent that statutory and regulatory preemption are different—under the Court's larger jurisprudence—difficulty may be expected in applying the same set of preemption rules to both areas."); Mendelson, *supra* note 2, at 739 ("When faced with an agency interpretation addressing a statute's preemptive effect, courts have trod unevenly in reconciling *Chevron* deference with the *Rice* presumption against preemption."); Nelson, *supra* note 8, at 232–33 ("Most commentators who write about preemption agree on at least one thing: Modern preemption jurisprudence is a muddle."); Sharkey, *supra* note 2, at 454 ("It is exceedingly difficult to demonstrate that any consistent principle or explanatory variable emerges from the Supreme Court's products liability preemption jurisprudence.").

54. *See, e.g.*, Wyeth v. Levine, 129 S. Ct. 1187, 1201 (2009) (citing *Skidmore* and reasoning that the weight accorded to an agency's preemption determination depends on its "thoroughness, consistency, and persuasiveness").

55. *See* Watters v. Wachovia Bank, N.A., 550 U.S. 1, 41 (2007) (Stevens, J., dissenting) ("Even if the [agency] did intend its regulation to pre-empt the state laws at issue here, it would still not merit *Chevron* deference. No case from this Court has ever applied such a deferential standard to an agency decision that could so easily disrupt the federal–state balance. . . . [U]nlike Congress, administrative agencies are clearly not designed to represent the interests of States, yet with relative ease they can promulgate comprehensive and detailed regulations that have broad pre-emption ramifications for state law. For that reason, when an agency purports to decide the scope of federal pre-emption, a healthy respect for state sovereignty calls for something less than *Chevron* deference." (citation omitted) (internal quotation marks omitted)).

Second, even in the not-entirely-certain category of cases in which the court deems "something less" than *Chevron* deference to be appropriate, it is unclear what precisely something less entails. Thus, in *Medtronic, Inc. v. Lohr*[56] the Court was "substantially informed by"[57] an agency's view of its regulations' preemptive scope, whereas in *Wyeth v. Levine*[58] the Court accorded deference only based on the "thoroughness, consistency, and persuasiveness"[59] of the agency's explanation, and in *Riegel v. Medtronic, Inc.*[60] the Court proceeded "[n]either accepting nor rejecting the proposition that [a] regulation can properly be consulted to determine [a] statute's meaning."[61] The Court's unwillingness or inability to articulate a clearly defined standard has invited heavy criticism and no shortage of proposals from commentators.[62] Until the Court charts a clear course, the doctrine remains in limbo.

Third, the Court's view on the presumption against preemption is severely fractured. In any given preemption case it is nearly impossible to predict whether the presumption will make an appearance.[63] Commentators have long criticized the Court's halfhearted and haphazard application of the doctrine.[64] Where it makes an appearance, a finding against preemption is sure to follow, but predicting its appearances is difficult at best.[65] The Court's recent three-way split in *Wyeth v. Levine* regarding the presumption's applicability illustrates the problem nicely.

---

56. 518 U.S. 470 (1996).

57. *Id.* at 495. Compare the *Medtronic* Court's language with that of *Geier v. American Honda Motor Co.*, 529 U.S. 861, 863 (2000), according "some weight" to an agency's conclusion that state law would stand as an obstacle to federal goals.

58. 129 S. Ct. 1187 (2009).

59. *Id.* at 1201 (citing *Skidmore v. Swift & Co.*, 323 U.S. 134, 140 (1944)).

60. 128 S. Ct. 999 (2008).

61. *Id.* at 1011.

62. *See supra* Parts I, II.E.

63. *Compare, e.g.*, Medtronic, Inc. v. Lohr, 518 U.S. 470, 484–85 (1996) (applying the presumption to interpret the Medical Device Act), *with* Riegel v. Medtronic, Inc., 128 S. Ct. 999 (2008) (failing even to mention the presumption when interpreting the same statute).

64. *See* Sharkey, *supra* note 2, at 458 ("Here, I join a veritable chorus of scholars pointing out the Court's haphazard application of the presumption. In the realm of products liability preemption, the presumption does yeoman's work in some cases while going AWOL altogether in others." (footnotes omitted)); Calvin Massey, *"Joltin' Joe Has Left and Gone Away": The Vanishing Presumption Against Preemption*, 66 ALB. L. REV. 759 (2003).

65. Sharkey, *supra* note 2, at 506 ("[W]here [the presumption] rears its head, its effect is seemingly outcome determinative.").

Although the *Wyeth* majority relied on the presumption as a "cornerstone" of its decision, the dissent countered that the presumption is irrelevant in the context of conflict preemption.[66] Concurring in judgment and splitting the Court a third way, Justice Thomas reserved the question of the applicability of the presumption, finding its resolution unnecessary given the clarity of the relevant statutes and regulations.[67] Given such diversity of views even among the Justices, it is unsurprising that commentators criticize the doctrine as an ad hoc rationalization lacking explanatory power.[68]

### B. Explaining the Court's Haphazard Approach

Commentators have rightly criticized the Court's inconsistent approach to *Chevron*–preemption questions. The cases present a haphazard jumble of noncommittal and ambiguous statements of selective deference to agency determinations of preemption. Echoing that vein of criticism, this Article began with a critique of the Court's inconsistency. Before moving on, it is useful to pause and reflect on the reasons underlying the Court's reluctance to articulate a clear standard. Several fundamental causes appear to animate the Court's jurisprudence, and an examination of these causes explains both why the Court has been reluctant to apply across-the-board *Chevron* deference to agency preemption determinations and, perhaps more interestingly, what an appropriate framework for *Chevron*-style deference might look like in the preemption context.

#### 1. Sidestepping the Danger of Conflicting-Canon Gridlock

At root, the Court's inconsistency stems from the clash between two competing interpretive canons. *Chevron* suggests deference to agency views, whereas the presumption against preemption counsels against preemption absent strong evidence of congressional intent. Where the two conflict— that is, where an agency views state law as an obstacle to congressional statutory objectives but Congress does not itself clearly state an intent to preempt—the canons pull in opposite directions, leaving the Court with no clear-cut answer. Faced with such a conflict, the Court could simply assert the superiority of one canon over the other and proceed to apply that canon as usual.[69] This would be the proper course were it obvious that the rationale underlying one canon or the other was simply inapplicable given the unique circumstances of agency preemption determinations. Both canons, however, retain at least some persuasive force. It is neither unreasonable to suppose that Congress would intend an expert agency to make some preemptive determinations nor to suppose that Congress would speak clearly if it intended a statute to have preemptive effect. Therefore, it would be a mistake to privilege one canon to a position of complete superiority over the other.

---

66. *Wyeth v. Levine*, 129 S. Ct. 1187, 1194–95 & n.3 (2009); *id.* at 1228–29 (Alito, J., dissenting).
67. *Id.* at 1208 n.2 (Thomas, J., concurring).
68. Sharkey, *supra* note 2, at 506.
69. This is the approach of those who advocate across-the-board *Chevron* deference, treating preemption cases no differently than other instances of agency interpretation, and also of those who argue that no deference at all is due to agency preemption determinations.

Recognizing this, the Court has avoided conclusively embracing or rejecting the *Chevron* doctrine's applicability to preemption questions.[70] Instead, the Court has sidestepped canon-conflict gridlock by digging beneath the canons and focusing directly on congressional intent. In *Wyeth*, for example, the Court began its analysis with the "cornerstone" principle that "the purpose of Congress is the ultimate touchstone in every preemption case."[71] Later, considering the preemptive effect of a preamble published with a Food and Drug Administration (FDA) regulation, the Court refused deference to the agency's view because it conflicted with the Court's interpretation of congressional intent.[72] This focus on congressional intent results from an intense consciousness of the Court's role as a protector of federalism and skepticism regarding the applicability of *Chevron*'s underlying rationale.[73] Preemption determinations implicate federalism values, the consideration of which is outside agencies' traditional realms of expertise.[74] So, rather than blindly defer to agency determinations where the *Chevron* rationale may be inapplicable, the Court is careful to ascertain whether Congress intended a preemptive result.

---

*See, e.g.*, Young, *supra* note 2, at 869–71 (arguing that any deference is inappropriate in preemption cases); *Leading Cases*, *supra* note 2, at 272 (advocating universal application of *Chevron*).

70.  *See supra* Part II.A.

71.  *Wyeth*, 129 S. Ct. at 1194–95 (majority opinion) (quoting Medtronic, Inc. v. Lohr, 518 U.S. 470, 485 (1996)).

72.  *Id.* at 1201.

73.  *Cf.* Mendelson, *supra* note 2, at 755–56 (attributing the Court's use of the presumption against preemption to a "reluctance to risk incidental statutory interference with federalism values and with state sovereignty" and "attaching substantive value to federalism goals"); *id.* at 779–91 (suggesting that agency determinations regarding preemption should not be accorded *Chevron* deference because agencies lack institutional competence to make such decisions).

74.  *Id.*

*2. Selective Use of the Presumption Against Preemption*

This focus on congressional intent helps to explain the Court's selective and seemingly haphazard application of the presumption against preemption.[75] Though the Court has expressed greater-than-average concern for congressional intent in the context of preemption, it still works within the traditional doctrinal framework. One element of that framework is the presumption against preemption. Rather than abandon the presumption in the face of canon conflict, the Court has, by selective application, converted the presumption into an important component of its intent-focused jurisprudence.

Although the presumption's influence has gradually waned over the last few decades, it has retained its force in cases of implied preemption, where congressional intent is least certain.[76] The principle's selective invocation allows it to serve as a thumb on the balance against preemption where the Court is least certain of congressional intent. Thus, the presumption was absent from *Riegel v. Medtronic, Inc.*, where an express preemption clause revealed an unmistakable intent to preempt,[77] but was invoked with force in *Wyeth*, where congressional intent was far less certain. Of course, the presumption is not invoked in every implied preemption case. As the *Wyeth* dissent points out, for instance, the presumption was notably absent in *Geier v. American Honda Motor Co.*[78] Even there, however, the presumption's use

---

    75.  *See supra* Part II.A.
    76.  *See* Altria Grp., Inc. v. Good, 129 S. Ct. 538, 556 (2008) (Thomas, J., dissenting) ("Since *Cipolone*, the Court's reliance on the presumption against pre-emption has waned in the express pre-emption context."). *Compare* Rowe v. N.H. Motor Transp. Ass'n., 552 U.S. 364 (2008), Riegel v. Medtronic, Inc., 128 S. Ct. 999 (2008), Watters v. Wachovia Bank, N.A., 550 U.S. 1 (2007), Engine Mfrs. Ass'n. v. S. Coast Air Quality Mgmt. Dist., 541 U.S. 246 (2004), Sprietsma v. Mercury Marine, 537 U.S. 51 (2002), Geier v. Am. Honda Motor Co., 529 U.S. 861 (2000), United States v. Locke, 529 U.S. 89 (2000), *and* Freightliner Corp. v. Myrick, 514 U.S. 280 (1995) (failing to apply the presumption against preemption in the presence of an express preemption clause), *with Wyeth*, 129 S. Ct. at 1194–95, Pharm. Research and Mfrs. of Am. v. Walsh, 538 U.S. 644, 666 (2003), Rush Prudential HMO, Inc. v. Moran, 536 U.S. 355, 365–66 (2002), Johnson v. Fankell, 520 U.S. 911, 918 (1997), California v. ARC Am. Corp., 490 U.S. 93, 101 (1989), *and* Hillsborough Cnty. v. Automated Med. Labs., Inc., 471 U.S. 707, 716 (1985) (applying the presumption in the context of implied preemption). *But see Altria Group*, 129 S. Ct. at 543–44 (majority opinion); Bates v. Dow Agrosciences LLC, 544 U.S. 431, 449 (2005); *Lohr*, 518 U.S. at 485; Cipollone v. Liggett Grp., Inc., 505 U.S. 504, 518 (1992) (applying the presumption in the context of express preemption).
    77.  128 S. Ct. at 1003.
    78.  *Wyeth*, 129 S. Ct. at 1228–29 (Alito, J., dissenting) ("[T]he *Geier* Court specifically rejected the argument (again made by the dissenters in that case) that the 'presumption against pre-emption' is relevant to the conflict pre-emption analysis. Rather than invoking

appears tied to congressional intent. *Geier* was ultimately decided on a theory of implied preemption,[79] but the statute at issue did contain an express preemption clause.[80] Congress had explicitly stated its intent to preempt *something*; the Court was simply faced with the question of whether it might implicitly have intended to preempt other aspects of state law as well. Thus, while there was a danger of overpreempting within an area already the target of some preemption, there was no danger of preempting an area of law where Congress had intended no preemption whatsoever.

In the absence of a congressionally expressed intent to preempt, the Court is hesitant to infringe on areas of traditional state sovereignty, and it demonstrates that concern by its selective use of the presumption against preemption. This insight explains why the Court's use of the principle appears haphazard at first glance. By selective application, the principle's invocation curtails the reach of agency power where the Court is concerned that congressional intent may be lacking.

### 3. *Varying Deference to Agency Determinations*

The Court's inconsistent standard of deference to agency determinations of preemption, like its inconsistent application of the presumption against preemption, appears to hinge on its concern for congressional intent and state sovereignty.[81] The Court reserves its most deferential language for cases where congressional intent to preempt is clear. Where, on the other hand, congressional intent is less certain, the Court either neglects to mention agency views or treats them as useful only to the extent persuasive. Thus in *Medtronic, Inc. v. Lohr*, the Court was "substantially informed by"[82] the agency's view, and in *Geier*, the agency's position was entitled to "some weight,"[83] but in *Wyeth*, where an express preemption clause was lacking, the Court treated the agency's view as merely one among many potentially

---

such a 'presumption,' the Court emphasized that it was applying 'ordinary,' 'longstanding,' and 'experience-proved principles of conflict pre-emption.'" (citations omitted)).

79. *Geier*, 529 U.S. at 866.

80. *Id.* at 867.

81. Deference to agency determinations and the presumption against preemption are two sides of the same interpretive coin, and so it is unsurprising that both doctrines' applicability in a given case depends on the same considerations. If the presumption against preemption is accorded its full weight as a traditional tool of statutory construction capable of resolving textual ambiguities, statutes would rarely, if ever, be found to contain the ambiguously preemptive language necessary for *Chevron* deference to apply. The Court can either give *Chevron* its full weight or give the presumption its full weight, but not both. *See* Mendelson, *supra* note 2, at 745–46.

82. *Id.* at 495.

83. *Geier*, 529 U.S. at 883.

persuasive authorities.[84] When certain of congressional intent to preempt, the Court appears willing to accord substantial weight to agency views, even on questions that implicate federalism.

*4. The* Chevron–Mead *Failure*

We are now in a position to understand the Court's unwillingness to define the relationship between the *Chevron* doctrine of deference and the presumption against preemption. Confronted with a clash between the canons, the Court sidesteps the gridlock by focusing on congressional intent. But it does not always sidestep in the same direction. In some cases, particularly express preemption cases where congressional intent to preempt is fairly clear, the Court is willing to rely heavily on agency determinations. In such cases the *Chevron* rationale of congressional delegation to superior agency expertise appears quite reasonable. Congress has an objective in mind that will require some preemption of state law, but rather than define the precise contours of that preemption, Congress delegates that decision to an expert agency. In other cases, where congressional intent is less clear, the Court is unwilling to defer to agency determinations. There the *Chevron* rationale is undercut because Congress has not clearly articulated an intent to preempt, and the agency is claiming power not only to define the scope of preemption but to determine whether there is to be any preemption at all. It is much less likely that Congress intended to delegate this greater power. So, focusing on intent, the Court selectively invokes the presumption against preemption or deference to agency determinations depending on which is a more accurate indicator of likely congressional intent in a given case.

This focus on intent explains why the Court has been unwilling to articulate a clear standard. Neither across-the-board *Chevron* deference, nor across-the-board *Skidmore* deference, nor even across-the-board nondeference would permit the Court to focus on congressional intent. The Court's jurisprudence suggests a desire to vary deference based on the presence or absence of congressional intent to delegate decisionmaking authority to the agency, and any across-the-board framework is, by definition, incapable of such variation.

Consider a rule of universal *Chevron* deference. As a number of scholars have noted, such a regime would risk errant intrusion into areas of traditional state sovereignty because agencies, while experts within certain congressionally delegated spheres, lack competence to balance power between the federal and state governments.[85] Although in some instances congressional delegation of narrow preemptive power to an agency would be quite reasonable,[86] such a delegation would be unthinkable (or at least highly unusual) in other contexts. Across-the-board *Chevron* deference would sweep up both sets of cases together, ignoring their significant differences.

---

84. Wyeth v. Levine, 129 S. Ct. 1187, 1201 (2009).

Even *Mead*'s modification of the *Chevron* doctrine to account for congressional intent to delegate fails to remedy the difficulty. *Mead*'s focus is on formality: "[E]xpress congressional authorization[] to engage in the process of rulemaking" is "a very good indicator of delegation meriting *Chevron* treatment."[87] Such formal rulemaking authority, while arguably[88] a useful indicator of intent to delegate generally, is a less useful indicator of intent to delegate *preemptive* authority.

In some circumstances, congressional silence regarding preemption may quite reasonably be viewed as an ambiguous gap into which an agency may insert its reasonable interpretation via its rulemaking power. Where Congress expressly indicates an intent to preempt all state law that poses an obstacle to a particular statutory objective, for instance, it intentionally leaves the scope of preemption vague.[89] Under such circumstances, the grant of rulemaking authority to the administering agency indicates a desire to have that gap filled by the agency. Where, however, Congress says nothing at all about preemption, it is much harder to read that silence, even if accompanied by rulemaking authority, as an implicit delegation of preemptive authority. Preemption is the sort of question about which silence is often not ambiguous, and so the presence or absence of rulemaking power says little about Congress's intent to delegate.

---

85. *See* Watters v. Wachovia Bank, N.A., 550 U.S. 1, 41 (2007) (Stevens, J., dissenting) (arguing that "state sovereignty calls for something less than *Chevron* deference" because administrative agencies' regulations have "broad pre-emption ramifications for state law" despite their role in representing only federal interests); Mendelson, *supra* note 2, at 779–91.

86. For instance, power to define the precise contours of an express preemption clause. *See infra* Part III.B discussing, as examples, the Medical Device Act and the Motor Vehicles Safety Act.

87. United States v. Mead Corp., 533 U.S. 218, 229 (2001).

88. *Mead*'s formulation requires both that an agency be granted official rulemaking authority by Congress and, further, that the agency exercise that authority in promulgating its resolution of the statutory ambiguity. *Id.* As Justice Scalia notes in dissent, the connection between a grant of rulemaking authority and intent to delegate is not itself particularly strong, and it is even harder to see why an agency should be required to exercise that authority when pronouncing its interpretation. If Congress intended to delegate authority it should make no difference how an agency makes its view known. *Id.* at 246 (Scalia, J., dissenting).

89. Express but ambiguously broad preemption of this sort is quite common. *See, e.g.*, Medtronic, Inc. v. Lohr, 518 U.S. 470 (1996) (considering the preemptive effect of the Medical Device Act's preemption clause).

Of course, *Mead* leaves open the possibility that factors other than force-of-law formality could guide the Court's analysis of congressional intent. Notice-and-comment rulemaking is neither necessary nor sufficient for deference to an agency's view.[90] At least in theory, then, the Court could avoid *Mead*'s force-of-law test and engage in naked examination of congressional intent to delegate. In practice, however, the *Mead* doctrine has been applied in a rule-like fashion. And furthermore, were the doctrine actually to devolve into a case-by-case search for congressional intent, *Chevron* would lose all utility as a bright-line rule, and all *Chevron* cases would be thrown into the same unpredictable chaos that currently grips the *Chevron*–preemption line. In short, all of Justice Scalia's worst fears would be realized.[91]

A rule of across-the-board *Skidmore* deference would suffer from similar defects. *Skidmore* deference initially presents itself as an appropriately middling alternative to a strict regime of *Chevron* deference or across-the-board nondeference. It avoids problems of agency incompetence to strike federal–state power balances by varying deference to agency interpretations depending on their "thoroughness, consistency, and persuasiveness."[92] A number of commentators have suggested such an approach,[93] and the Court itself at times appears inclined toward such a rule.[94] If deference varies based on agency competence, what could possibly go wrong? The Court's love–hate relationship with the doctrine hints at the answer.

Although the Court sometimes applies *Skidmore*-like deference to agency determinations of preemption, other times it hints at something more—"*Skidmore*-with-bite" it might be called.[95] The Court applies deference in this way because it recognizes that not all agency preemption decisions are created equal. Sometimes Congress may very well intend to delegate limited preemptive authority to agencies. To apply the traditional *Skidmore* analysis of agency persuasiveness would be to second-guess Congress's decision to delegate. Something more akin to *Chevron* deference is really in order, and so the Court applies *Skidmore*-with-bite.

---

90. *See Mead*, 533 U.S. at 230–31 (majority opinion) ("That said, and as significant as notice-and-comment is in pointing to *Chevron* authority, the want of that procedure here does not decide the case, for we have sometimes found reasons for *Chevron* deference even when no such administrative formality was required and none was afforded.").

91. *See id.* at 246–61 (Scalia, J., dissenting) (opposing vigorously *Mead*'s exception to the *Chevron* rule).

92. Wyeth v. Levine, 129 S. Ct. 1187, 1201 (2009) (citing Skidmore v. Swift & Co., 323 U.S. 134, 140 (1944))

93. *See supra* notes 2–5.

94. *See, e.g.*, *Wyeth*, 129 S. Ct. at 1201 (noting that deference depends on the agency's "thoroughness, consistency, and persuasiveness" and citing *Skidmore*); Riegel v. Medtronic, Inc., 128 S. Ct. 999, 1009 (2008) (finding it unnecessary to consider the agency's view because the statute itself was clear, but noting that had it considered the agency's position, "mere *Skidmore* deference would seemingly be at issue").

95. *See, e.g.*, Medtronic, Inc. v. Lohr, 518 U.S. 470, 495 (1996) (where the Court was "substantially informed by" the agency's view).

A pure rule of *Skidmore* deference cannot accommodate this need. Although an across-the-board *Skidmore* regime would eliminate the possibility of errant preemption due to agency inexpertise where Congress does not intend to delegate, it cannot provide full *Chevron*-style deference in those situations where congressional intent so requires. In such cases a pure *Skidmore* regime would take interpretive power given to agencies by Congress and subject it to the determination of a judiciary applying the notoriously malleable *Skidmore* test.[96]

The unpredictable nature of the Court's *Chevron*–preemption jurisprudence ultimately stems from its focus on congressional intent. Any across-the-board framework would be too inflexible to produce the varying levels of deference that are appropriate in preemption cases, and so the Court has avoided committing to a particular approach. In failing to provide a consistent framework, however, the Court has sacrificed predictability. Under *Chevron*, regulated parties can rely on courts to uphold agency determinations. But in the preemption context, all certainty has been lost. The analysis has devolved into a case-by-case assessment of congressional intent.

### III. UNMUDDLING THE COURT'S *CHEVRON*–PREEMPTION JURISPRUDENCE

The Court's inconsistency in *Chevron*–preemption cases stems from the perceived need to undertake a case-by-case search for congressional intent. Congressional intent to delegate preemptive authority varies widely from case to case, and rather than sweep up all cases into the same across-the-board framework of deference, the Court has applied deference selectively in some cases but not others. The Court has carved out an area of law where the *Chevron* rule of deference is not universally applicable and, within that area, replaced the *Chevron* rule with a policy of case-by-case weighing of congressional intent.

---

96. *See* Kristin E. Hickman & Matthew D. Krueger, *In Search of the Modern* Skidmore *Standard*, 107 COLUM. L. REV. 1235, 1237 (2007) ("All agree that *Skidmore* is less deferential than *Chevron*, but how much less and in what way remain open questions."). A rule of across-the-board nondeference would suffer all of the same flaws as a pure *Skidmore* regime. A rule of general nondeference would prevent errant agency determinations of preemption, but it would ignore the will of Congress where Congress decided to delegate limited preemptive authority to expert agencies.

This imprecise and unpredictable approach has attracted a great deal of academic criticism, and rightly so.[97] Regardless of which side of the rules–standards debate one takes, the Court's jurisprudence is entirely unsatisfactory. Not only has the Court failed to produce a clear rule, it has failed even to produce a consistent fuzzy standard. Given this inconsistency, the uniformity of an across-the-board *Skidmore* or *Chevron* rule is an attractive alternative. Bright-line rules are by nature over- and underinclusive, and imperfect accuracy is a necessary sacrifice to obtain predictability.

But the Court has rightly resisted such approaches. The stakes are unusually high in preemption cases, and the Court should ensure that it is Congress initiating any and all preemptive lawmaking.[98] Relying on an across-the-board assumption of *Skidmore* or *Chevron* deference disrespects actual congressional intent because, in the context of preemption, neither a presumption of full deference nor a presumption of limited deference is appropriate across all cases. Sometimes Congress intends to delegate preemptive authority to agencies, and in such cases *Chevron* deference is warranted and mere *Skidmore* deference is overly intrusive. Other times, when Congress does not intend to delegate preemptive authority, even *Skidmore* deference is inappropriate. Across-the-board solutions fail to account for Congress's intent to delegate because preemption is unique. Because of preemption's importance and agencies' relative lack of expertise, statutory ambiguity does not always (but sometimes does) imply congressional intent to delegate.

### *A. A Bright-Line Alternative to the Court's Haphazardry*

Faced with a choice between rule-like certainty and respect for congressional intent, the Court has sided with congressional intent. But it need not make this either–or decision. A bright-line rule carefully crafted to account for congressional intent would avoid both horns of the dilemma, providing much needed certainty while still respecting *Chevron*'s variable applicability to preemption questions.

---

97. *See supra* Part II.A.
98. So great is Justice Thomas's concern for state sovereignty in preemption cases that he would even go so far as to abandon the Court's obstacle preemption jurisprudence altogether. Instead he would find preemption only where Congress has clearly spoken or it would be impossible to comply with both state and federal requirements. *See Wyeth*, 129 S. Ct. at 1204–08 (Thomas, J., concurring).

The Court's recent decisions point to a possible solution. The Court, guided in large part by its concern to protect federalism and its respect for congressional intent, has balanced the *Chevron* doctrine and presumption against preemption quite differently depending on the presence or absence of an express preemption clause.[99] When considering a statute with an express preemption clause, the Court is much less likely to invoke the presumption against preemption and much more likely to defer to preemptive agency determinations. Alternatively, when considering a statute that lacks an express preemption clause, the Court is less deferential to agency determinations and more likely to apply the presumption against preemption.

This pattern is unsurprising given the Court's focus on intent. Congress speaks most clearly when it utilizes an express preemption clause. But the Court should rely on the presence or absence of an express preemption clause as much more than a strong indicator of congressional intent. It should replace its fuzzy, intent-focused analysis with a bright-line rule of full *Chevron* deference to agency interpretations when a statute contains an express preemption clause and nondeference in the absence of such a clause. Such a rule would provide the certainty of a bright-line rule while still respecting congressional intent. Furthermore, it would reconcile the purposes underlying both the *Chevron* doctrine and the presumption against preemption and relieve tension between the doctrines by allocating to each determinative power within an exclusive category of cases.

### B. *Respecting Congressional Intent and State Sovereignty*

A rule that varies deference based on the presence or absence of an express preemption clause closely follows congressional intent to delegate. Where Congress has expressly stated an intent to preempt state law but has left the statute ambiguous as to the scope of that preemption, it is reasonable to infer an intent to delegate authority to the administering agency to determine the appropriate scope of preemption. For instance, in *Medtronic, Inc. v. Lohr* the Court considered the preemptive effect of the Medical Devices Act of 1976, which reads in pertinent part: "[N]o State . . . may establish or continue in effect with respect to a [medical] device . . . any requirement—(1) which is different from, or in addition to, any requirement applicable under this chapter . . . and (2) which relates to the safety or effectiveness of the device."[100] Such language is clearly intended to preempt *something*, but the scope of preemption is left unelaborated. Given its intentional vagueness, it is hard to read this language as anything but an intentional delegation of preemptive authority to the agency responsible for administering the statute. Indeed, the Court itself all but concluded as much:

> The FDA regulations interpreting the scope of [the statute's] pre-emptive effect support the [plaintiffs'] view, and our interpretation of the pre-emption statute is substantially informed by those regulations. . . . Because the FDA is the federal agency to which Congress has delegated its authority to implement the provisions of the Act, the agency is uniquely qualified to determine whether a particular form of state law "stands as an obstacle to the accomplishment and execution of the full purposes and objectives of Congress," and, therefore, whether it should be pre-empted.[101]

---

   99. *See supra* Part II.B.2–3.
   100. Medical Device Act of 1976 § 521, 21 U.S.C. § 360k (2006).

Where Congress has clearly expressed an intent to preempt state law but left the scope of that preemption vague, courts can reasonably presume an intent to delegate and full *Chevron*-style deference is in order. Doubts of agency competency and canons respecting traditional areas of state sovereignty have no place where Congress has expressed an intent to displace state law and affirmed its faith in the administrative agency's competency to handle the task.

Full *Chevron* deference is occasionally in order even in somewhat less obvious cases. In *Geier v. American Honda Motor Co.*,[102] for instance, the Court considered the preemptive effect of the National Traffic and Motor Vehicle Safety Act,[103] which includes both an express preemption clause and a

---

101. *Medtronic*, 518 U.S. at 495–96 (footnote omitted) (quoting Hines v. Davidowitz, 312 U.S. 52, 67 (1941)). Note, though, that the Court refrained from granting full *Chevron*-style deference to the agency's view. Its determinations only "substantially informed" the Court. *Id.* Justice Breyer, concurring in part and in the judgment, was even more explicit regarding the deference due to an agency under such circumstances:

> [T]he MDA's [Medical Device Act's] pre-emption provision is highly ambiguous. That provision makes clear that federal requirements may pre-empt state requirements, but it says next to nothing about just when, where, or how they may do so. . . . Thus, Congress must have intended that courts look elsewhere for help as to just which federal requirements pre-empt just which state requirements, as well as just how they might do so . . . .
>
> [T]his Court has previously suggested that, in the absence of a clear congressional command as to pre-emption, courts may infer that the relevant administrative agency possesses a degree of leeway to determine which rules, regulations, or other administrative actions will have pre-emptive effect."

*Id.* at 505–06. (Breyer, J., concurring in part and concurring in the judgment) (citing, *inter alia*, Chevron U.S.A. Inc. v. Natural Res. Def. Council, Inc., 467 U.S. 837 (1984)).

102. 529 U.S. 861 (2000).

103. National Traffic and Motor Vehicle Safety Act of 1966, 15 U.S.C. §§ 1381–1431 (1988).

savings clause explicitly preserving state common law.[104] In that case, after finding the plaintiff's negligence claim not expressly preempted, the Court considered the possibility of obstacle preemption.[105] When considering obstacle preemption in the shadow of an express preemption clause, the Court should be deferential to reasonable agency views.[106] In such cases Congress has already clearly expressed an intent to preempt some state law, and the agency is an expert within its sphere of authority. An inference of intent to delegate in such circumstances is at least as reasonable as the inference[107] of congressional intent to preempt that is implicit in any finding of obstacle preemption.[108]

In both types of cases the concerns that typically militate against the application of *Chevron* deference to agency preemption determinations are absent. Where Congress is silent on the question of preemption, Courts are rightly hesitant to defer to agencies' preemptive decisions. The Constitution empowers Congress, not executive agencies, to preempt state

---

104. The Act's preemption clause reads as follows:

    Whenever a Federal motor vehicle safety standard established under this subchapter is in effect, no State or political subdivision of a State shall have any authority either to establish, or to continue in effect, with respect to any motor vehicle or item of motor vehicle equipment any safety standard applicable to the same aspect of performance of such vehicle or item of equipment which is not identical to the Federal standard.

*Id.* § 1392(d). The effect of that clause was tempered, however, by the Act's savings clause: "Compliance with [a federal safety standard] does not exempt any person from any liability under common law." *Id.* § 1397(k).

105. *Geier*, 529 U.S. at 869–70 (concluding that neither a savings clause nor an express preemption clause prohibits the ordinary workings of obstacle preemption). The dissenters in *Geier* disputed the majority's consideration of obstacle preemption after an express preemption clause has already been found to be inapplicable. *Id.* at 900 n.16 (Stevens, J., dissenting). Without taking a position on that debate, this Article simply assumes that such analysis is sometimes appropriate and considers the level of deference due.

106. In *Geier* the agency expressed its interpretation only in its brief to the Court. *See id.* at 911. In light of *Mead*, an agency would now likely need to express its view more formally to merit deference.

107. The obstacle preemption doctrine assumes that Congress would have intended to displace state law that poses an obstacle to federal objectives. *See* Nelson, *supra* note 8, at 228–29 ("So-called 'obstacle preemption' potentially covers not only cases in which state and federal law contradict each other, but also all other cases in which courts think that the effects of state law will hinder accomplishment of the purposes behind federal law.").

108. In recent years commentators have advanced strong arguments that the Court should curtail or eliminate its use of implied preemption doctrines. I do not intend, here, to take a position on the advisability of obstacle preemption as a general matter. I merely note that as long as the Court continues to embrace the doctrine, it should apply the appropriate level of deference to agency views. *See id.* at 229 n.16; Wyeth v. Levine, 129 S. Ct. 1187, 1204–08 (2009) (Thomas, J., concurring) (criticizing the Court's obstacle preemption jurisprudence).

law,[109] and agency decisions are not subject to the political and procedural safeguards that protect states against preemptive congressional action. Agencies lack the direct democratic accountability of Congress, and their actions are not restrained by the elaborate procedural requirements of the federal legislative process.[110] These arguments have less force, however, where Congress has made a decision to preempt state law and the only question is the scope of that preemption. In such cases, the question of preemption has already been subjected to the rigors of the democratic process. Where Congress has spoken clearly but imprecisely in favor of preemption, Courts should defer to reasonable agency determinations of preemptive scope.

Where, on the other hand, Congress has said nothing at all about an intent to preempt state law, it is much less likely that Congress intended to delegate preemptive authority to the agency. The possibility of illegitimate federal incursion into spheres of state sovereignty looms large. The Court's recent decision in *Wyeth v. Levine* provides a useful example. There, the Court considered the preemptive effect of the Food, Drug, and Cosmetic Act (FDCA) in relation to the plaintiff's common-law negligence claim.[111] The defendant, Wyeth, argued that even though the statute contains no express preemption clause, it should nonetheless be read to preempt the plaintiff's claim because state negligence law stands as an obstacle to Congress's purpose of creating a uniform federal regulatory regime.[112] In support of its preemption argument, Wyeth cited a preamble to a 2006 FDA regulation in which the agency declared that the FDCA should be read to establish both a floor and a ceiling for drug labeling, preempting conflicting State labeling laws.[113]

---

109. U.S. CONST. art. VI, cl. 2 ("This Constitution, and the Laws of the United States . . . shall be the supreme Law of the Land; and the Judges in every State shall be bound thereby, any Thing in the Constitution or Laws of any State to the Contrary notwithstanding."); *see* Bradford R. Clark, *Process-Based Preemption*, *in* PREEMPTION CHOICE: THE THEORY, LAW, AND REALITY OF FEDERALISM'S CORE QUESTION 192, 192–93 (William W. Buzbee ed., 2009) (noting the Supremacy Clause's negative implication that state law governs in the absence of "supreme Law" and that the Senate holds an absolute veto over the adoption of every source of law identified by the Supremacy Clause as supreme law).

110. *See* Young, *supra* note 2, at 869–70 ("The states have no direct role in the 'composition and selection' of federal administrative agencies, and much of the point of such agencies is to be more efficient lawmakers than Congress. Agency action thus evades both the political and the procedural safeguards of federalism." (footnotes omitted)). *See generally* Larry D. Kramer, *Putting the Politics Back into the Political Safeguards of Federalism*, 100 COLUM. L. REV. 215 (2000); Herbert Wechsler, *The Political Safeguards of Federalism: The Rôle of the States in the Composition and Selection of the National Government*, 54 COLUM. L. REV. 543 (1954).

111. *Wyeth*, 129 S. Ct. at 1191.

112. *Id.* at 1193–94.

Unlike both the Medical Device Act and the Motor Vehicles Safety Act, the FDCA contains no express preemption clause. In the absence of an express preemption clause, the Court was rightly critical[114] of the agency's interpretation:

> If Congress thought state-law suits posed an obstacle to its objectives, it surely would have enacted an express pre-emption provision at some point during the FDCA's 70-year history. But despite its 1976 enactment of an express pre-emption provision for medical devices, Congress has not enacted such a provision for prescription drugs. Its silence on the issue, coupled with its certain awareness of the prevalence of state tort litigation, is powerful evidence that Congress did not intend FDA oversight to be the exclusive means of ensuring drug safety and effectiveness.[115]

Where Congress has not spoken regarding preemption, courts should be skeptical of agency views. Indeed, they should be more than skeptical; they should be completely nondeferential. Where Congress is silent, its intent to delegate preemptive authority to an agency is least plausible, and the possibility of unauthorized federal intrusion is at its highest. Even *Skidmore* deference would unacceptably undervalue federalism and flout congressional intent. As applied, the *Skidmore* standard tends to be highly deferential to agency views.[116] And even if applied in the less deferential fashion that the Court sometimes employs,[117] *Skidmore* would interfere with Congress's ability to call the preemptive shots.[118]

In the absence of an express preemption clause assuring congressional intent to preempt, agency views should be accorded no special weight.

---

113. *Id.* at 1200; Requirements on Content and Format of Labeling for Human Prescription Drug and Biological Products, 71 Fed. Reg. 3922, 3934–35 (Jan. 24, 2006) (codified at 21 C.F.R. pts. 201, 314, 601 (2011)).

114. The Court did, however, grant the agency's position *Skidmore*-like deference based on its "thoroughness, consistency, and persuasiveness." *Wyeth*, 129 S. Ct. at 1201.

115. *Id.* at 1200.

116. *See* Hickman & Krueger, *supra* note 96, at 1280–81 ("[A]nalysis of 106 identified *Skidmore* applications in the federal courts of appeals demonstrates that, in a strong majority of cases, the *Skidmore* doctrine represents a bona fide standard of review, rather than merely an excuse for reviewing courts to follow their own interpretive preferences. Additionally, the evidence shows that *Skidmore* review is highly deferential—less so than *Chevron*, but still weighted heavily in favor of government agencies over their challengers.").

117. *See id.* at 1252–53 (reading *Christensen v. Harris County*, 529 U.S. 576, 587 (2000), to apply a particularly nondeferential version of the *Skidmore* standard).

118. *See* Young, *supra* note 2, at 891 ("What any version of *Skidmore* appears to rule out, moreover, is any sort of presumption *against* the agency's interpretation, such as that which the *Rice* presumption against preemption would impose if the agency's interpretation displaced state law.").

Where, however, Congress has expressed its intent to preempt clearly but has only vaguely defined the statute's preemptive scope, agencies should be accorded full *Chevron*-style deference.[119] This approach would provide a much more rule-like and predictable framework than the Court's current case-by-case analysis while still preserving state sovereignty and respecting congressional intent.

### C. *Consistency with* Chevron*'s Rationale and Reconciliation of Conflicting Canons*

Not only would a rule of variable deference predicated on express preemption respect state sovereignty and congressional intent, it would also reconcile the *Chevron* rule of deference with the presumption against preemption by allocating to each an independent sphere of influence. The various across-the-board proposals for *Chevron*–preemption deference resolve the tension between the two canons in one of two ways. First, blending together the two offsetting concerns and producing a happy medium, several commentators have suggested *Skidmore* deference as the universal solution in *Chevron*–preemption cases.[120] This approach successfully reconciles the canons but at the expense of ignoring their full force.

In cases where Congress almost certainly did not intend to delegate preemptive authority to an agency, undermining *Chevron*'s rationale, *Skidmore* applies even though nondeference and application of the presumption against preemption would be a more appropriate solution. In cases where Congress did intend to delegate authority and *Chevron*'s rationale does apply, *Skidmore* is applied in its stead. In both instances a

---

119. With this proposed rule I do not intend to suggest that the presence or absence of an express preemption clause qualitatively distinguishes two distinct types of congressional action for only one of which deference is due to agency views. The qualitative distinction between statutes that do and do not demand deference hinges solely on congressional intent. I suggest merely that the presence or absence of an express preemption clause is a predictable, text-based proxy for such intent, reliance on which would be superior to the Court's current approach. Like any bright-line rule, it is both over- and underinclusive. Congress may sometimes intend delegation and fail to include an express preemption clause, or vice versa.

Rules of this sort suffer another common limitation as well: they are definitionally incapable of accommodating fine distinctions in legislative intent. If, for instance, without expressly indicating so, Congress intended a middling standard of judicial review between *Chevron* and nondeference, or delegation of authority only to decide a particular preemptive question, the framework would be unable to produce the intended result. A text-based rule of variable deference tracks congressional intent more closely than do across-the-board rules, but as it remains a bright-line rule, it still suffers their familiar shortcomings, though to a lesser degree.

120. *See supra* note 5 and accompanying text.

single one-size-fits-all standard is applied, ignoring the fit of the canons' rationales.

Second, another set of commentators, rather than blend the conflicting canons together into a happy medium, pits them in a fight to the death, maximizing the victor across all cases.[121] Such solutions, which claim either that the interests underlying the presumption against preemption trump *Chevron* or vice versa, ignore an entire set of cases where the competing canon's rationale is actually the stronger of the arguments. Across-the-board nondeference, for instance, would withhold deference even where Congress has included an express preemption clause and seems to have intended an agency to resolve the statute's ambiguous preemptive scope.

Neither across-the-board maximization nor across-the-board compromise accurately reflects *Chevron*'s waxing and waning force. Deference contingent on the presence of an express preemption clause, however, accounts for the variable applicability of *Chevron* across cases. Consider three factors often cited as weighing against *Chevron* deference in preemption cases: federalism, agency inexpertness, and agency self-aggrandizement.[122]

First, because agencies are not politically accountable directly to the states and their procedure for making law is much freer than Congress's, some argue that agency preemption poses a special danger to state sovereignty.[123] *Chevron* deference would be inappropriate, they argue, because it circumvents the traditional safeguards of federalism. But when deference is made contingent on an express preemption clause, this danger is significantly lessened. An agency unilaterally undertaking a decision as to whether there is to be any preemption at all would pose a potentially serious danger to state sovereignty. Only the federal legislature is authorized to exercise such power. But where Congress, a democratically elected and procedurally burdened body has already made a decision to preempt state law, and an agency is tasked with determining only the scope of that preemption, the danger of illegitimate federal incursion is significantly reduced. The agency is working within a limited sphere of delegated authority, and most importantly, *Congress* has already made the decision to displace obstacular state law.

Second, relative agency inexpertness is also commonly advanced against *Chevron* deference to preemptive agency interpretations. Agencies are frequently criticized for their "tunnel vision." They tend to focus intensely on particular programmatic objectives to the detriment of broader, system-wide goals.[124] Although agencies have great familiarity with the statute they are responsible for administering, they are relatively inexperienced in allocating power between the federal and state governments.[125] Given their limited competence to address the constitutional dimension of preemption questions, the argument goes, full *Chevron*-style deference to agency views would be inappropriate in the preemption context.[126]

---

    121. *See supra* notes 3–4 and accompanying text.
    122. *See* Mendelson, *supra* note 2, at 779–97; Young, *supra* note 2, at 869–71, 890–91.
    123. *See, e.g.*, Young, *supra* note 2, at 869–71.

These criticisms are well founded. Congress is far better suited than agencies to strike the proper federal–state balance of power. But where Congress has already struck the balance in favor of preemption and the only question is preemptive scope, agency competency is no longer such an issue. Agencies are quite well suited to answer questions of this second sort. Their skill is in the particularization of broad policy objectives, and it is precisely this skill that is called into use when Congress ambiguously calls for preemption of any state law that poses an obstacle to its goals. Of course, there may be a bit of overpreemption at the margins. Agencies are most skilled only in determining where uniformity would further federal objectives, not whether a certain level of nonuniformity might nonetheless be desirable in light of federalism values. But such overpreemption would occur only where Congress had already expressed a desire to preempt at least some state law. Furthermore, one might argue that agency federalism expertise is not even relevant to the question of *Chevron* deference once Congress has spoken clearly in favor of preemption. At that point Congress has already devalued federalism in a particular context, and agencies are not responsible for revaluing it in determining preemptive scope.[127]

Third, the danger of agency self-aggrandizement is sometimes cited as weighing against the application of *Chevron* deference. An agency could, by reading an ambiguous statute to preempt state law, increase its own importance by making itself the exclusive regulator. "[A]llowing agencies to define the scope of their own authority runs headlong into the venerable constitutional principle that 'foxes should not guard henhouses.'"[128] Where federal preemption is at stake, *Chevron* deference would inappropriately permit an agency to define the limits of its own power.[129] Where Congress has already expressly preempted some indeterminate amount of state law, however, this argument carries somewhat less weight. An agency might face a temptation to expand its power slightly at the edges, but those edges are defined by Congress, which makes the initial call for uniformity. When Congress has already gone so far as to aggrandize an agency with an express preemption clause, only a relatively insignificant amount of additional self-aggrandizement is possible.

---

124. *See* Mendelson, *supra* note 2, at 780–81 (noting as an example environmental agencies' tendency to focus on eliminating the last bit of risk presented by known hazards rather than addressing more significant risks).

125. *See* Merrill, *supra* note 2, at 755–56.

126. *See* Mendelson, *supra* note 2, at 779–85.

127. *Id.* at 790–91.

128. Young, *supra* note 2, at 889 (quoting Cass R. Sunstein, *Interpreting Statutes in the Regulatory State*, 103 HARV. L. REV. 405, 446 (1989)).

129. *See* Mendelson, *supra* note 2, at 795–96 n.243.

Unlike across-the-board deference, a rule of variable deference made contingent on an express preemption clause accounts for *Chevron*'s varying applicability in preemption cases. Congress is unlikely to delegate preemptive authority in a way that endangers federalism or risks agency self-aggrandizement. Neither is it likely to delegate questions about which agencies lack expertise. For these reasons *Chevron*'s rationale of delegation to agency expertise is undermined in implied preemption cases. Recognizing that *Chevron* deference would be inappropriate and that Congress likely did not intend to delegate preemptive authority, courts should grant no special deference to agency views. Instead they should apply the presumption against preemption along with all other tools in their interpretive arsenal and treat the agency's view as just one coequal voice among many.

This nondeference, however, should not be applied across the board. Just as the presumption against preemption has its exclusive sphere of influence where *Chevron*'s rationale is undercut, *Chevron* should enjoy its sphere of influence as well. Where Congress has spoken through an express preemption clause, concerns of federalism, expertise, and self-aggrandizement are outweighed by Congress's expressed intent and *Chevron*'s rationale of agency delegation. Ambiguity regarding the scope of express preemption is just the sort of question that Congress might reasonably intend to be decided by an expert agency. Courts should recognize this and defer to reasonable agency interpretations of ambiguous preemptive scope.

Only a rule of variable deference can account for the waxing and waning force of *Chevron*'s underlying rationale. Such a rule would relieve the tension between *Chevron* and the presumption against preemption by assigning to each a sphere[130] of independent influence. Further, it would

---

130. Note that this framework's provision for independent spheres of influence also resolves, at least in the agency preemption context, the protracted disagreement among the Justices regarding the applicability of the presumption against preemption where a statute contains an express preemption clause. Does the presumption continue to hold force in express preemption cases, militating in favor of a narrow construction, or does the presumption lose force in the face of conclusive preemptive intent? *See* Cipollone v. Liggett Grp., Inc., 505 U.S. 504, 545 (1992) (Scalia, J., dissenting) (disputing the majority's use of provide a more rule-like determinate of deference than does the Court's current case-by-case investigation of congressional intent.

### D. *A Narrow and a Broad Framework for Implementation*

A rule of variable deference contingent on express preemption could be grafted onto the Court's current *Chevron*–preemption jurisprudence in at least two ways: one conservative, the other slightly more daring.

#### 1. *A* Mead*-Like Delegation Rule*

First, the Court could craft a new rule modeled after *Mead*'s force-of-law test. The *Mead* Court recognized congressional authorization to engage in rulemaking to be "a very good indicator of delegation meriting *Chevron* treatment."[131] Although the *Mead* test itself is not well suited to the preemption context,[132] the Court could simply announce a new preemption-specific interpretive principle: the presence of an express preemption clause indicates delegation meriting *Chevron* deference, and the absence of such a clause indicates nondelegation.[133]

This approach would meld neatly with the Court's existing *Chevron*-preemption jurisprudence. The Court has consistently applied something like *Skidmore* deference, but it has varied its use of the presumption against preemption and tailored its application of *Skidmore* deference on a spectrum ranging from *Skidmore*-with-bite to minimal deference. Because, under *Mead*, even a finding of nondelegation results in *Skidmore* deference,[134] the

---

the presumption against preemption to narrowly interpret a statute's express preemption clause).

This debate would be rendered largely moot if full *Chevron* deference were granted to an agency's interpretation of an express preemption clause. A court's only interpretive role would be to assess the reasonableness of the agency's construction—an analysis in which the presumption against preemption would be unlikely to play a part. The presumption is useful as a tiebreaking rule of thumb and does not offer the degree of certainty that would be required to support a finding that an agency's interpretation was unreasonable.

131. United States v. Mead Corp., 533 U.S. 218, 229 (2001).
132. *See supra* Part II.B.4.
133. This formulation, unlike the *Mead* test, presents a condition that is both necessary and sufficient for *Chevron* deference. Without a bright-line safe harbor for agency deference, the Court's jurisprudence could fall back into its current pattern of unpredictable case-by-case decisionmaking. *Compare Mead*, 533 U.S. at 230–31 (leaving open the possibility of *Chevron* deference even in the absence of formal agency action), *with id.* at 245–46 (Scalia, J., dissenting) (interpreting the *Mead* majority to announce a more rule-like safe-harbor rule and fearing that the majority's *Chevron* exception making and invocation of the notoriously indeterminate *Skidmore* standard will lead to "protracted confusion").
134. *Id.* at 234 (majority opinion) ("To agree with the Court of Appeals that Customs ruling letters do not fall within *Chevron* is not, however, to place them outside the pale of any Court could apply a similar range of deference under the proposed framework. Currently the Court reserves its most deferential version of *Skidmore* deference for express preemption cases and its least deferential version of *Skidmore* for implied preemption cases.[135] The only change that the proposed framework would demand is a transition from *Skidmore*-with-bite to full *Chevron*-style deference. Everything else would remain the same. The Court's decisions would be more predictable because they would be based on a concrete rule, and they would be slightly more deferential to agency views where Congress provides an express preemption clause, but on the whole the Court's decisions would look much as they now do.

The one flaw of this framework, of course, is that it would grant *Skidmore* deference where nondeference would be more appropriate. While the rule would recognize *Chevron*'s full force in express preemption cases, it would ignore the force of the presumption against preemption in implied preemption cases. Agency decisions regarding preemptive scope, made in the shadow of an express preemption clause, suffer from none of the defects that militate against the application of *Chevron* to preemption questions. But where Congress has not spoken, these defects weigh heavily against any deference at all to agency views. The risk of agency self-aggrandizement is at its highest, agencies lack expertise to consider federal–state constitutional issues, and it is unlikely that Congress intended to delegate such power. Under such circumstances nondeference and application of the presumption against preemption and other traditional tools of interpretation are the more fitting solutions.

### 2. A Chevron-*Based Delegation Rule*

Second, and more daring, the Court could graft a new preemption rule onto *Chevron* itself. *Chevron*'s traditional formulation for review of agency statutory interpretations requires a two-step inquiry: first a court examines the statute to determine whether Congress has "directly spoken to the precise question at issue";[136] second, if the statutory language is indeed ambiguous, the question is whether the agency's interpretation is a

---

deference whatever. *Chevron* did nothing to eliminate *Skidmore*'s holding that an agency's interpretation may merit some deference whatever its form, given the 'specialized experience and broader investigations and information' available to the agency, and given the value of uniformity in its administrative and judicial understandings of what a national law requires." (quoting Skidmore v. Swift & Co., 323 U.S. 134, 139–40 (1944)).

135. *See supra* Part II.B.2–3.
136. Chevron U.S.A. Inc. v. Natural Res. Def. Council, Inc., 467 U.S. 837, 842 (1984).

reasonable construction of the statute. If both requirements are met, a court must defer to the agency's interpretation.[137]

As we have seen, a rule that makes deference contingent on an express preemption clause allocates nonoverlapping spheres of applicability to the *Chevron* principle and the presumption against preemption. *Chevron*'s rationale wins out in express preemption cases, and the presumption against preemption's rationale wins out in implied preemption cases. It is reasonable to presume congressional intent to delegate preemptive authority where Congress supplies an express preemption clause, and it is reasonable to presume against such delegation where Congress supplies no such clause. If these canons are viewed not only as "winning out" within their respective spheres but as doing so in a very particular way— definitively answering a question of congressional intent—the presumption against preemption that applies absent an express preemption clause can be seen as resolving the *Chevron* inquiry at step one and thus eliminating any need for deference at *Chevron* step two. Put simply, if absent an express preemption clause the presumption against preemption resolves any statutory ambiguity as to delegation of preemptive authority,[138] *Chevron* deference is applicable in express preemption cases and inapplicable in implied preemption cases.

*U.S. Telecom Ass'n v. FCC*[139] provides a usefully analogous example. There the Court of Appeals for the District of Columbia Circuit considered whether the Telecommunications Act of 1996[140] permits the Federal Communications Commission (FCC) to subdelegate a portion of its authority under the Act to state commissions.[141] Because the statute did not explicitly foreclose the possibility of subdelegation, the FCC argued that its interpretation of the statute to permit subdelegation should be entitled to deference under *Chevron*.[142] The court forcefully rejected this argument at

---

137. *Id.* at 843–44.
138. Under this formulation the presumption serves not to answer the question of preemption itself but, rather, the question of congressional delegation. It would more accurately be called a "presumption against delegation of preemptive authority" than a "presumption against preemption." The presumption resolves, at *Chevron* step one, only the question of delegation. Otherwise, it would be impossible for a court ever to find in favor of nonexpress varieties of preemption. If the statute contained an express preemption clause, *Chevron* deference would be in order, and if it contained no such clause, it would be found unambiguously nonpreemptive. I do not advocate such a use of the presumption. Rather, at *Chevron* step one, the presumption merely rules out congressional delegation of preemptive authority, not the possibility of preemption itself.
139. 359 F.3d 554 (D.C. Cir. 2004).
140. Pub. L. No. 104-104, 110 Stat. 56, codified at 47 U.S.C. § 151 (2006).
141. *See U.S. Telecom Ass'n*, 359 F.3d at 564–65.
142. *Id.* at 565.

*Chevron* step one, finding the statute not even to be ambiguous on the question of subdelegation:

> The Commission's plea for *Chevron* deference is unavailing. A general delegation of decision-making authority to a federal administrative agency does *not*, in the ordinary course of things, include the power to subdelegate that authority beyond federal subordinates. It is clear here that Congress has not delegated to the FCC the authority to subdelegate to outside parties. The statutory "silence" simply leaves that lack of authority untouched. In other words, the failure of Congress to use "Thou Shalt Not" language doesn't create a statutory ambiguity of the sort that triggers *Chevron* deference.[143]

Congress need not create a laundry list of every action that an agency is prohibited from undertaking in order to avoid *Chevron*-inducing ambiguity. "Were courts to *presume* a delegation of power absent an express *withholding* of such power, agencies would enjoy virtually limitless hegemony, a result plainly out of keeping with *Chevron* and quite likely with the Constitution as well."[144] In some contexts congressional silence is best read not as an ambiguity but as a clear intent not to delegate a particular power.

Just as a statute that fails to use the magic words "thou shalt not subdelegate to nonagency bodies" is not truly ambiguous as to subdelegative power, a statute that includes no express preemption clause is not ambiguous as to Congress's desire to delegate preemptive authority to an agency.[145] Federal preemption, like extraagency subdelegation, is an unusual administrative power, and when it is left unmentioned, it is reasonable to read a statute not to confer it. By not speaking, Congress actually speaks quite clearly.

### 3. Contrasting the Approaches

A rule of variable deference enforced at *Chevron* step one is superior to a *Mead*-like rule for several reasons: it offers greater predictability, it more accurately reflects the canon's modes of operation, and it respects the exclusive spheres of *Chevron* and the presumption against preemption. First, and most obviously, a *Chevron*-based rule would provide a much more predictable standard of deference in *Chevron* preemption cases. Both *Mead* and *Skidmore* provide fairly indeterminate rules of decision, and the incorporation of both into a single rule is a recipe for uncertainty.[146] A *Mead*-like rule could be rendered much more predictable if, contrary to *Mead* itself, the presence of an express preemption clause was made both a necessary and sufficient condition for delegation, but *Skidmore*'s indeterminacy is irredeemable.

---

143. *Id.* at 566.
144. Ry. Labor Executives' Ass'n v. Nat'l Mediation Bd., 29 F.3d 655, 671 (D.C. Cir. 1994) (emphasis in original).
145. As noted *supra* note 138, the statute may, of course, still be ambiguous as to preemptive effect. A court may find a statute clearly not to delegate preemptive authority while still finding, and judicially resolving in favor of implied preemption, an ambiguity regarding the implied preemptive effect of the statutory scheme.

Second, a *Chevron*-based rule would more accurately reflect the canons' modes of operation. A *Mead*-like rule would be predicated on the notion that an express preemption clause of ambiguous scope demonstrates an intent to delegate preemptive authority and the lack thereof demonstrates intent not to delegate. But the lack of an indeterminate express preemption clause is more appropriately seen not as nondelegation but as nonambiguity. A statute that includes no express preemption clause unambiguously intends no delegation of preemptive authority, just as a statute making no mention of extraagency subdelegation unambiguously confers no such power.

Third, a rule enforced at *Chevron* step one would, unlike a *Mead*-based rule, fully respect the power of the presumption against preemption. Under *Mead*, an agency is entitled to *Skidmore* deference even if it is found ineligible for full *Chevron* deference.[147] The application of *Skidmore* deference in implied preemption cases (where *Chevron*'s rationale is entirely inapplicable) would needlessly blunt the effect of the presumption against preemption within its sphere of applicability. *Skidmore* as applied tends to be quite deferential and its application could counterbalance or completely outweigh the presumption against preemption.[148] A *Chevron*-based rule, by contrast, would recognize the full force of both canons. In express preemption cases an agency would be entitled to full *Chevron*-style deference and in implied preemption cases it would receive no deference at all, giving the presumption against preemption free range.

Considered collectively, these factors weigh in favor of a *Chevron*-based rule rather than a *Mead*-based one. The first promotes clarity and respects the scope and rationales of the canons, whereas the second provides somewhat less clarity and deviates from its foundational rationales. Ultimately, however, the choice of *Mead* or *Chevron* is much less significant than the project as a whole. *Chevron* is best, but either would be a significant improvement.

---

146. *See* United States v. Mead Corp., 533 U.S. 218, 246–250 (2001) (Scalia, J., dissenting) (opposing the *Mead* exception to *Chevron* partially on the ground that it would lead to uncertainty in the lower courts). *See generally* Hickman & Krueger, *supra* note 96, at 1311–20 (presenting an empirical analysis of *Skidmore*'s varied application across 106 cases).

147. *See Mead*, 533 U.S. at 234 (majority opinion) ("To agree with the Court of Appeals that Customs ruling letters do not fall within *Chevron* is not, however, to place them outside the pale of any deference whatever. *Chevron* did nothing to eliminate *Skidmore*'s holding that an agency's interpretation may merit some deference whatever its form, given the specialized experience and broader investigations and information available to the agency, and given the value of uniformity in its administrative and judicial understandings of what a national law requires." (citation omitted) (internal quotation marks omitted)).

148. *See supra* note 116 and accompanying text.

CONCLUSION

This analysis begins the project of unraveling the Court's tangled knot of *Chevron*–preemption jurisprudence. The Court's haphazard decisionmaking stems from its high regard for congressional intent when considering questions that affect the federal–state balance of power. The rule of *Chevron* deference and the presumption against preemption provide conflicting measures of congressional intent, and, rather than universalize one principle or the other, the Court has applied a middling standard of *Skidmore* deference on a sliding scale—sometimes quite deferentially and sometimes almost nondeferentially—depending on its case-by-case analysis of congressional intent.

Critics who propose uniform, across-the-board deference in all cases recognize the flaw in the Court's approach—its unpredictability—but they fail to recognize its merits: respect for congressional intent, state sovereignty, and *Chevron*'s underlying rationale. In express preemption cases, the Court does not need to enforce federalism values through the presumption against preemption because Congress has spoken clearly in favor of displacing state law. And if the scope of preemption is ambiguous, *Chevron*'s presumption of delegation through ambiguity to agency expertise is quite reasonable. On the other hand, where Congress has not spoken clearly through an express preemption clause, *Chevron*'s rationale is particularly weak. Nondeference and application of the presumption against preemption are in order. A regime of uniform deference across all cases is unable to account for *Chevron*'s waxing and waning force.

Instead, the Court should adopt a rule of variable deference that accords full *Chevron*-style deference to agency interpretations of ambiguously broad express preemption clauses and withholds deference altogether where Congress is silent regarding preemption. Such a rule would recognize the factors that underlie the Court's unpredictable case-by-case approach— respect for state sovereignty and congressional intent—while providing the rule-like certainty demanded by the Court's critics.